\documentclass[aps,twocolumn,superscriptaddress]{revtex4-2}

\usepackage{hyperref}
\hypersetup{colorlinks=true, citecolor=blue, urlcolor=blue, linkcolor=blue}

\usepackage{orcidlink}
\usepackage{graphicx}
\usepackage{xcolor}
\usepackage{dcolumn}
\usepackage{bm}
\usepackage{amsmath}
\usepackage{amssymb}
\usepackage{mathtools}

\newcommand{\tr}{\text{Tr}}

\begin{document}
\title{Critical Scaling of the Quantum Wasserstein Distance}

\author{Gonzalo Camacho\orcidlink{0000-0001-6900-8850}}\email{gonzalo.camacho@dlr.de}
 \affiliation{%
 Department High-Performance Computing, Institute of Software Technology, German Aerospace Center (DLR), 51147 Cologne, Germany
}
\author{Benedikt Fauseweh\orcidlink{0000-0002-4861-7101}}\email{benedikt.fauseweh@tu-dortmund.de}
 \affiliation{%
 Department High-Performance Computing, Institute of Software Technology, German Aerospace Center (DLR), 51147 Cologne, Germany
}%
\affiliation{%
Department of Physics, TU Dortmund University, Otto-Hahn-Str. 4, 44227, Dortmund, Germany
}

\begin{abstract}
Distinguishing quantum states with minimal sampling overhead is of fundamental importance to teach quantum data to an algorithm. Recently, the quantum Wasserstein distance emerged from the theory of quantum optimal transport as a promising tool in this context. Here we show on general grounds that the quantum Wasserstein distance between two ground states of a quantum critical system exhibits critical scaling.  We demonstrate this explicitly using known closed analytical expressions for the magnetic correlations in the transverse field Ising model, to numerically extract the critical exponents for the distance close to the quantum critical point, confirming our analytical derivation. Our results have implications for learning of ground states of quantum critical phases of matter.
\end{abstract}

\date{\today}

\maketitle

Given two different probability distributions, finding the optimal mass transport between the two, {\color{black}{that is, the most efficient way of transforming one distribution to another,} }posses in general a challenging mathematical problem~\cite{Ambrosio2008} with recent applications in signal processing, data science and machine learning \cite{Kolouri2017,Wang2013,arjovsky2017wassersteingan,MAL-073}. In the quantum realm, the concept of optimal mass transport between distributions is extended to any two quantum states~\cite{KarolZyczkowski_1998,Biane2001,Carlen2014}, with their corresponding density matrices taking the role of probability distributions, whereby a \emph{quantum distance} between the two can be defined. When defining such distance, it is often desired that, contrary to more conventional measures like the fidelity or the relative entropy~\cite{Braunstein1994,Nielsen2010}, the considered metric is not unitarily invariant. 

A prominent example of a quantum distance is the quantum Wasserstein distance~\cite{Golse2016,Golse2018,Caglioti2020,DePalma2021a,DePalma2021,Geher2023,DePalma2023,Toth2023,Bunth2024}, which has gathered considerable attention as an optimal transport map between arbitrary quantum states. It defines a metric that is robust against local perturbations, with the property of being non-maximal between states having orthogonal supports. Such distinguishability metrics are useful in reducing the number of copies required for performing state estimation~\cite{Aaronson2007,DePalma2021}, identifying optimal algorithms for learning quantum data~\cite{Kiani2022,Anshu2024,Rouze2024,Rouze2024quantum,Zhang2024,Herr_2021,pmlr-v247-caro24a, Cao_2024}, describing optimal transport in nonequilibrium quantum thermodynamics~\cite{PhysRevLett.130.107101,PhysRevX.13.011013,VanVu2021,VanVu2023}, improving variational quantum algorithms~\cite{DePalma2023prxq,Zoratti2023}, estimating bounds for quantum error mitigation
\cite{PhysRevLett.131.210602}, and finding optimal tomography of unitary quantum circuits~\cite{Li2022} and parametrized quantum states~\cite{schreiber2024tomographyparametrizedquantumstates}. Despite many applications, fundamental properties of the Wasserstein distance are not yet explored, leaving the door open to develop connections between seemingly distant research fields.

From the quantum information perspective, it is of interest to find the role of the quantum Wasserstein distance in the theory and detection of entanglement~\cite{Horodecki2008,Friis2019}, for instance, by exploring its connection to different entanglement measures, witnesses, or alternative distances between quantum states. It has been shown that when the quantum Wasserstein distance reduces to the self-distance of a quantum state (i.e. the two states become the same), it is in direct relation with the Wigner-Yanase information~\cite{Wigner1963,DePalma2021a} if the optimization is carried out over general states. Recently, a direct relation between the self-distance of a quantum state and a multipartite entanglement witness, the quantum Fisher information (QFI), has been noted~\cite{Toth2023}. 

The relation between the quantum Wasserstein distance and the QFI suggests an immediate connection with the field of many-body quantum systems hosting quantum critical points, where the presence of multipartite entanglement can be witnessed experimentally through dynamic susceptibilities that are directly related to the QFI~\cite{Hauke2016,PhysRevB.108.184302,PhysRevB.98.134303,Wang_2014,camacho2024observingdynamicallocalizationtrappedion}. In turn, the relation implies that the limiting case of the quantum Wasserstein distance as a self-distance for quantum states is a measurable quantity for the detection and classification of quantum phase transitions~\cite{Sachdev2011}. An intriguing direction of fundamental relevance is therefore to explore the quantum Wasserstein distance in many-body systems known to host quantum phase transitions. In particular, characterizing universal scaling laws~\cite{Cardy1996} in the vicinity of such critical points might have important implications in the search of optimized learning protocols for many-body quantum states~\cite{Rouze2024,Rouze2024quantum,Cao_2024}, but also in characterizing the scaling behavior of these systems away from alternative more conventional metrics~\cite{Venuti2007}. Understanding the critical properties of the quantum Wasserstein distance could thus lead to lower measurement requirements for characterizing quantum phases of matter or to develop improved quantum machine learning algorithms \cite{Huang2020}.

In this article, we present results on the critical behavior of the quantum Wasserstein distance between two many-body ground states, and demonstrate this property of the distance in a paradigmatic model of a one-dimensional spin chain hosting nearest-neighbour interactions, the transverse field Ising model (TFIM). 

We follow the definitions from Ref.~\cite{Toth2023}. Given any two quantum states $\rho,\sigma$ living on a Hilbert space $\mathcal{H}$, the quantum Wasserstein distance of order 2 $D(\rho,\sigma)^2$ between the states is obtained by finding the coupling $\rho_{12}\in \mathcal{H}\otimes\mathcal{H}$ satisfying:
\begin{eqnarray}\label{eq:wasserstein_distance}
2D(\rho,\sigma)^2 &=&\text{min}_{\rho_{12}}\text{Tr}\left[(O\otimes I - I\otimes O)^2\rho_{12}\right],\nonumber\\
\tr_2(\rho_{12})&=& \rho, \tr_1(\rho_{12})=\sigma,
\end{eqnarray}
where $O$ is an Hermitian operator of an $N$-qubit system, and $\text{Tr}_1(\rho_{12})$ refers to the partial trace over the leftmost subspace in $\mathcal{H}\otimes \mathcal{H}$. 

We consider an extensive operator in the $N$-qubit system $O=\sum_{n=1}^N O_n$.
If at least one of the states $\rho,\sigma$ is pure, the Wasserstein distance has the simpler form~\cite{Toth2023}:
\begin{eqnarray}\label{eq:wd_def}
D(\rho,\sigma)^2 &=&\frac{1}{2}\langle O^2\rangle_\rho + \frac{1}{2}\langle O^2\rangle_\sigma - \langle O\rangle_\rho\langle O\rangle_\sigma, 
\end{eqnarray}
with $\langle O\rangle_\rho = \text{Tr}( O \rho)$. When both states become equal to each other $\rho=\sigma$, this corresponds to the QFI $F_Q\left[\rho,O\right]$ of the state,
\begin{eqnarray}\label{eq:self_distance}
D(\rho,\rho)^2=\langle O^2\rangle_\rho-\langle O\rangle_\rho^2=(\Delta O)^2=\frac{F_Q\left[\rho,O\right]}{4}.
\end{eqnarray}
A modified version of Eq.~\eqref{eq:wd_def} which is a true metric~\cite{Bunth2024} consists on subtracting the contributions from the self-distance Eq.~\eqref{eq:self_distance}. We refrain from using the modified version since the self-distance will also present critical scaling.

In the following we assume a parameterized Hamiltonian $H(g)$ on a lattice $\Lambda$ with lattice spacing $a=1$. {\color{black}We assume periodic boundary conditions in what follows. }At $g = g_c$ the system exhibits a quantum critical point. For $g < g_c$ the system is in its ordered phase with order parameter $O$. We consider two ground states $\rho$ and $\sigma$ with parameters $g_\rho - g_c = \tilde{g}_\rho < 0$ and $ g_\sigma - g_c = \tilde{g}_\sigma > 0$. Close to the critical point $|\tilde{g}_\rho| \ll 1$ and $|\tilde{g}_\sigma| \ll 1$ hold. {\color{black} In this case correlation functions of the form $\langle O_j O_{j+n} \rangle \propto n^{-\eta}$ decay algebracially with distance $n$ up to the correlation length $\xi(g_\rho)$ and $\xi(g_\sigma)$ respectively, or system size $L$, depending on which is larger. In particular, the exponent $\eta$ is related to the scaling dimension of the operators $O_j$. The contributions from the order parameter are $L^2 O^2\sim L^2 (-\tilde{g}_\rho)^{2\beta}$, with $\beta$ being the scaling exponent.} We use this to approximate the expression for the quantum Wasserstein distance by integrals in two cases.  

\emph{Assumption 1 (Finite size)}: We consider the case in which the lattice spacing is much smaller than the linear system size $L$ but the correlation length is even larger, $ 1\ll L \ll \xi(g_\sigma)$ and $ 1\ll L\ll \xi(g_\rho)$. Then the terms in Eq.~\eqref{eq:wd_def} can be expanded as follows~\cite{Cardy1996,Hauke2016}
\begin{eqnarray}\label{eq:main_eq_1}
\langle {O}^2\rangle_\rho &=& C_\rho +L^2 O^2(g_\rho) + L B_\rho \int_0^{L} \frac{1}{r^\eta} \mathrm{d}r 
\nonumber \\
&\approx&  C_\rho+ L^2 A_\rho \left(-\tilde{g}_\rho\right)^{2\beta} + B_\rho L^{2-\eta} ,  \\
\langle {O}^2\rangle_\sigma &=& C_\sigma+ L B_\sigma  \int_0^{L}  \frac{1}{r^\eta} \mathrm{d}r  
\nonumber \\
 &\approx& C_\sigma + B_\sigma L^{2-\eta},
\end{eqnarray}
where $A_\rho, B_\rho, B_\sigma$ are non-universal constants from the approximation of the sum as an integral with the correlation exponent $\eta$ and order parameter exponent $\beta$, and we have defined the contributions from local terms $C_{\rho/\sigma}=L\sum_n \langle O_n^2\rangle_{\rho/\sigma}$. The leading contribution is given by $L^2 A_\rho \left(-\tilde{g}_\rho\right)^{2\beta}$, as the ground state $\rho$ is skewed towards a ground state with finite order parameter and $\sigma$ is not, leading to
\begin{eqnarray}\label{eq:main_eq}
D(\rho,\sigma)^2 &\approx & L^{2}A_\rho \left(-\tilde{g}_\rho\right)^{2\beta}  + L^{2-\eta} \left( B_\rho + B_\sigma \right)\nonumber\\
&+& \left(C_\rho + C_\sigma\right).
\end{eqnarray}

\emph{Assumption 2 (Thermodynamic limit)}: Here we consider the case $L  \gg \xi(g_\sigma) \gg 1$ and $L  \gg \xi(g_\rho) \gg 1$ . Then the terms in Eq.~\eqref{eq:wd_def} can be expanded as follows
\begin{eqnarray}\label{eq:main_eq_prev}
\langle {O}^2\rangle_\rho &=& C_\rho +L^2 O^2(g_\rho) + L B_\rho \int_0^{\xi(\tilde{g}_\rho)} \frac{1}{r^\eta} \mathrm{d}r + \mathcal{O} \left( e^{-\frac{L}{\xi(\tilde{g}_\rho)}} \right) \nonumber \\
&\approx&  C_\rho+ L^2 A_\rho \left(-\tilde{g}_\rho\right)^{2\beta} + L B_\rho \xi(\tilde{g}_\rho)^{1-\eta},  \\
\langle {O}^2\rangle_\sigma &=& C_\sigma+ L B_\sigma  \int_0^{\xi(\tilde{g}_\sigma)}  \frac{1}{r^\eta} \mathrm{d}r  + \mathcal{O} \left( e^{-\frac{L}{\xi(\tilde{g}_\sigma)}} \right) \nonumber \\
 &\approx& C_\sigma + L B_\sigma \xi(\tilde{g}_\sigma)^{1-\eta}.
\end{eqnarray}
The subleading contributions are given by $ L B_\rho \xi(\tilde{g}_\rho)^{1-\eta}$ and $L B_\sigma \xi(\tilde{g}_\sigma)^{1-\eta}$. The correlation length scales as $\xi(\tilde{g}_\sigma)  \sim \tilde{g}_\sigma^{-\nu}$ and $\xi(\tilde{g}_\rho)  \sim \left(-\tilde{g}_\rho\right)^{-\nu'}$ respectively; absorbing a factor of $1/2$ in all constants and scaling by $L^2$, we obtain
\begin{eqnarray}\label{eq:main_eq_2}
\frac{D(\rho,\sigma)^2}{L^2} &\approx & A_\rho \left(-\tilde{g}_\rho\right)^{2\beta}  + \frac{1}{L^2}\left(C_\rho + C_\sigma\right)\nonumber\\
&+& \frac{1}{L}\left( B_\rho \left(-\tilde{g}_\rho\right)^{\eta \nu' - \nu'} + B_\sigma \tilde{g}_\sigma^{\eta \nu - \nu}  \right)
\end{eqnarray}
for the universal scaling of the quantum Wasserstein distance between ground states.

In the following we demonstrate this scaling using exact expressions for the Transverse Field Ising Model (TFIM):
\begin{eqnarray}
\mathcal{H}_{\text{TFIM}}=-J\sum_j \sigma_j^x\sigma_{j+1}^x -h\sum_j \sigma_j^z,
\end{eqnarray}
where $\sigma_j^x,\sigma_j^z$ represent the Pauli $x$ and $z$ matrices at site $j$ of the chain, respectively. The model is exactly solvable~\cite{Lieb1961,Pfeuty1970}. We define the parameter $g=h/J$. The model has critical behavior at $g=1$, where there is a quantum phase transition from an ordered $(g<1)$ phase into a disordered $(g>1)$ phase. The order parameter capturing the critical behavior in the model is the transverse field magnetization, given by $M_x=\sum_{i=1}^L \sigma_i^x$. {\color{black}{From now on, we take $O=M_x$ in the above expressions.} } 

The correlations $\langle \sigma_i^x \sigma_{i+n}^x\rangle$ are obtained using the exact expressions given in Ref.~\cite{Pfeuty1970}; these correlations depend on integrals of the form
\begin{eqnarray}
L(n)=\frac{1}{\pi}\int_0^\pi dk\frac{\cos(nk)}{\sqrt{1+\frac{1}{g^{2}}+\frac{2}{g}\cos(k) }}.
\end{eqnarray}
We employed  arbitrary floating point precission Python library "mpmath"~\cite{mpmath} to compute the integrals. We use the exact result from Pfeuty~\cite{Pfeuty1970} in the limit $L\to\infty$ for the magnetization density:
\begin{eqnarray}\label{eq:magne_av}
\lim_{L\to\infty}\frac{\langle M_x\rangle_\rho}{L}=\begin{cases}\left(1-g_\rho^2\right)^{1/8},\hspace{5pt} g_\rho \leq 1,\\
0,\hspace{10pt} g_\rho > 1. 
\end{cases}
\end{eqnarray}
In what follows, Eq.~\eqref{eq:magne_av} is assumed regardless of the system size $L$. {\color{black} We identify the scaling exponent $2\beta=1/4$. From Ref.~\cite{Pfeuty1970}, the correlations $\langle \sigma_i^x \sigma_{i+n}^x \rangle \propto n^{-1/4}$, which determines the correlation function exponent $\eta=1/4$. }.

\begin{figure}[!t]
\centering
\includegraphics[scale=0.6]{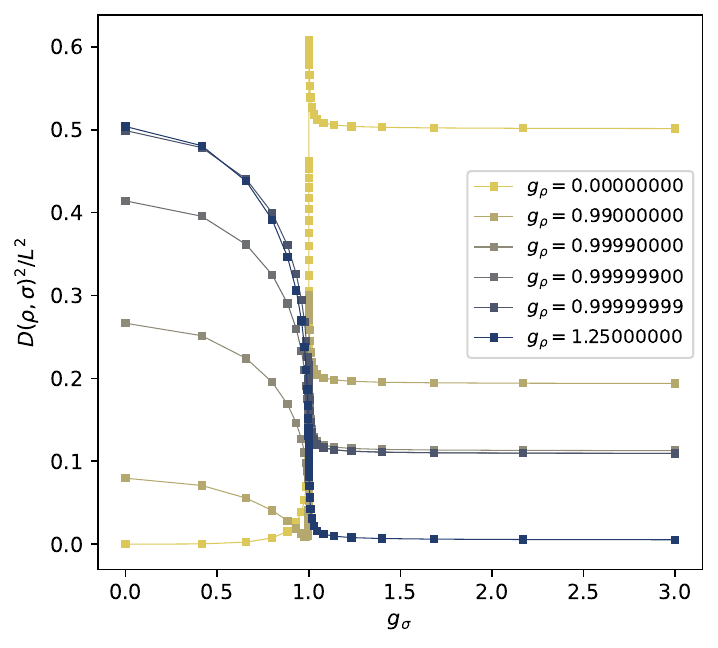}
\caption{The Wasserstein distance in Eq.~\eqref{eq:wd_def} scaled by $L^2$, as a function of $g_\sigma$ for different values of $g_\rho$, for a system size $L=500$. Regions of magnetic order and disorder are identified by the non-analytic behavior of the distance close to the quantum critical point.}
\label{fig:fig1}
\end{figure}

The Wasserstein distance between two ground states of the TFIM given by Eq.~\eqref{eq:wd_def} is represented in Fig.~\ref{fig:fig1} as a function of $g_\sigma$, for different values of $g_\rho$. Close to the quantum critical point, there are contributions from both $\langle M_x^2\rangle$ and $\langle M_x\rangle^2$; however, exactly at $g=1$ correlations due to $\langle M_x^2\rangle$ dominate, whereas in the ordered phase, $\langle M_x\rangle$ dominates the distance between the states. Note that for $g_\rho<1$, the distance between both ground states is almost always finite, stabilizing when $g_\sigma>1$. For $g_\rho=0,g_\sigma>1$ contributions due to $\langle M_x^2\rangle_\sigma$ and $\langle M_x\rangle_\sigma$ vanish in Eq.~\eqref{eq:wd_def}, leaving contributions from $\langle M_x^2\rangle_\rho$ only. From Ref.~\cite{Pfeuty1970}, we find $D(\rho,\sigma)^2\sim \frac{L\langle M_x\rangle_\rho}{2}$, hence $D(\rho,\sigma)^2/L^2\sim \frac{1}{2}$ in this limit, as shown in Fig.~\ref{fig:fig1}. Signatures of critical behavior are visible when approaching the quantum critical point, where $D(\rho,\sigma)^2/L^2$ develops non-analytic behavior even though both states are different. This shows that the quantum Wasserstein distance can be employed as a proxy to capture the critical behavior of the system.   

\begin{figure}[!t]
\centering
\includegraphics[scale=0.6]{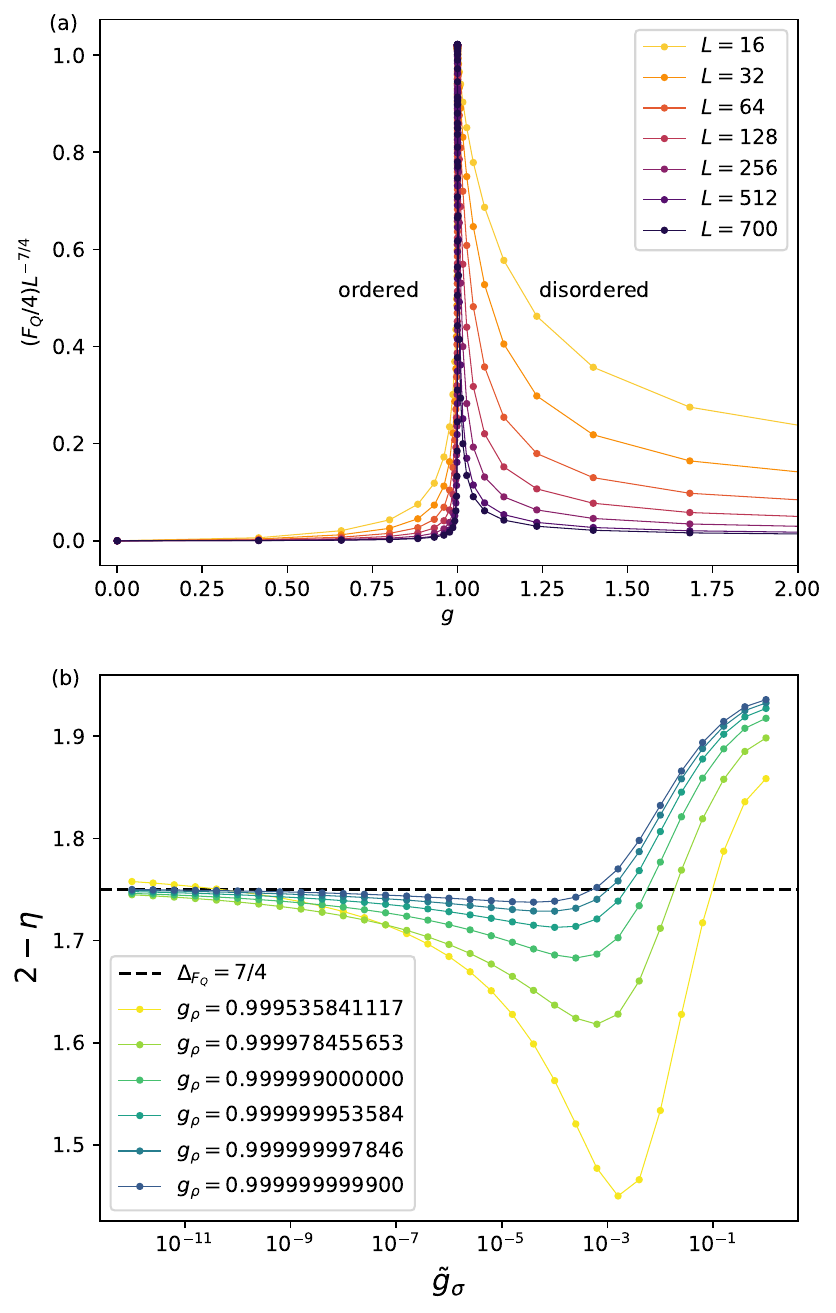}
\caption{(a) The QFI (self-distance) scaled by $L^{7/4}$, for different system sizes $L$. Close to the quantum critical point $g=1$, the QFI develops a narrow peak separating the two phases of the model when approaching the thermodynamic limit $L\to\infty$, in accordance with the results of Ref.~\cite{Hauke2016}. (b) The critical exponent for the quantum Wasserstein distance in the TFIM, when the two quantum states $\rho,\sigma$ are close to the critical point $g=1$, as a function of $\tilde{g}_\sigma=|g_\sigma - 1|$. In the limit $g_\rho,g_\sigma\to 1$, we obtain the scaling exponent from Ref.~\cite{Hauke2016}. We note that in regions away from the critical point, the scaling with $L$ might be ill-defined under Assumption 1. The different system sizes used to extract the exponents are $L=20,40,80,120,150,200,250,300,350,400,450,500,600,700$}
\label{fig:fig2}
\end{figure}

We begin the scaling analysis by looking at systems under Assumption 1. Based on Eq.~\eqref{eq:main_eq_1}, we explore the self-distance defined in Eq.~\eqref{eq:self_distance} for the TFIM, which is directly related to the QFI. The QFI for the TFIM is known to exhibit critical behavior around the quantum critical point~\cite{Sun2010, Liu2017, Hauke2016} with:
\begin{eqnarray}\label{eq:exp_qfi}
F_Q[\rho, M_x]\sim L^{\Delta_{F_Q}},  \hspace{5pt}\Delta_{F_Q}=\frac{7}{4}.
\end{eqnarray}
We observe that under Assumption 1, this scaling exponent is recovered analytically from Eq.~\eqref{eq:main_eq_1}, and identify $\Delta_{F_Q}=2-\eta$ iff $\rho=\sigma$. In Fig.~\ref{fig:fig2}(a), we represent the QFI scaled by $L^{7/4}$ for different system sizes $L$, as a function of the parameter $g$. This critical scaling confirms the result from Ref.~\cite{Hauke2016}, with the QFI developing a spike at the critical point. In Fig.~\ref{fig:fig2}(b), we represent the scaling of the distance as a function of $\tilde{g}_\sigma$ for different states $\rho,\sigma$ to the left and right vicinity of the critical point, respectively. We stress that in the limit considered here, values of $g_\rho, g_\sigma$ result in large correlation lengths $\xi\gg L$, thus holding under Assumption 1. The results show the scaling exponent $2-\eta$ for quantum states being close to the critical point; the limit $\xi\to\infty$ corresponds to having both states exactly critical, in which case we recover $\Delta_{F_Q}$ as the scaling exponent.

\begin{figure}[!t]
\centering
\includegraphics[scale=0.62]{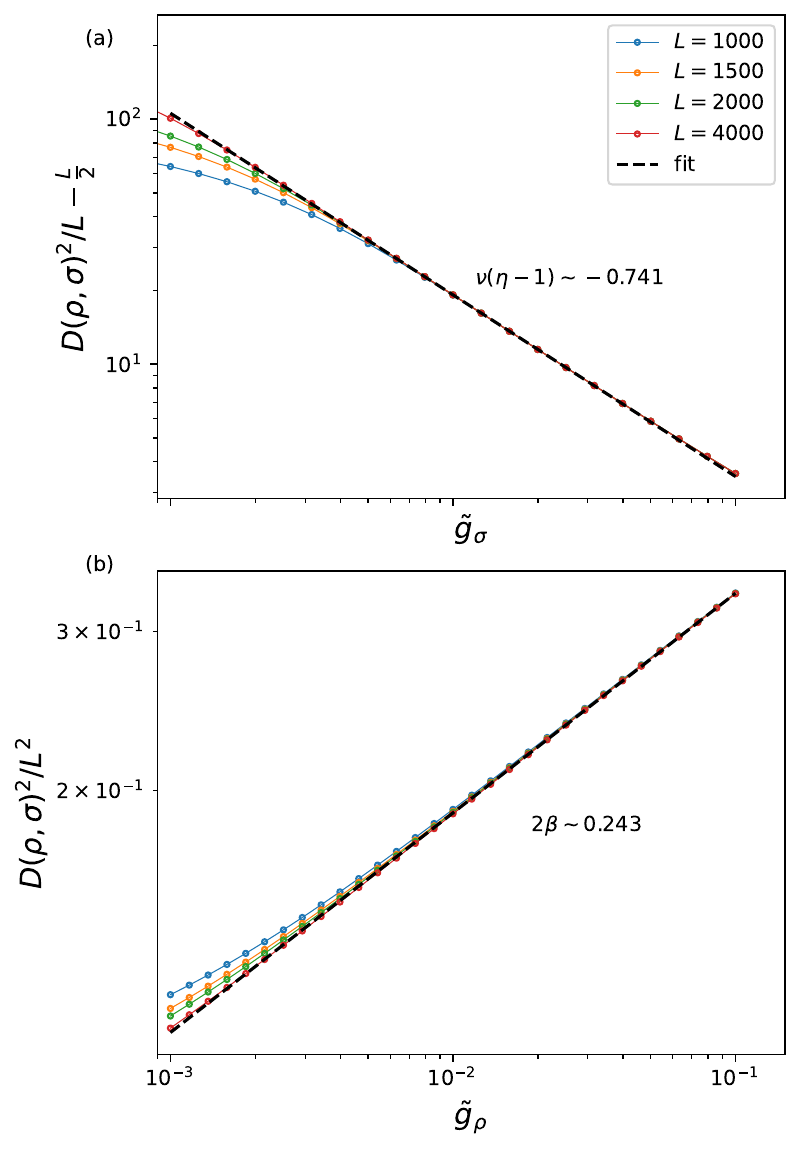}
\caption{(a) The power-law behavior for the sub-leading contribution in Eq.~\eqref{eq:main_eq}, for $g_\rho=0$ and the state $\sigma$ being close to the critical point, for different system sizes $L$. Note that since the contribution is subleading, we have scaled by a factor of $L$ after subtracting the analytic value of $A_\rho=\frac{1}{2}$. Note the logarithmic scale on both axes. The critical exponent is extracted numerically for the largest $L$, showing good agreement with the predicted analytic result $\nu(\eta-1)=-3/4$. (b) Power-law behavior of the leading contribution in Eq.~\eqref{eq:main_eq} when $g_\sigma=10$. The critical exponent is obtained numerically, in close agreement with the predicted analytic value for the TFIM $2\beta=\frac{1}{4}$. }
\label{fig:fig3}
\end{figure}

We consider now the system under Assumption 2, for the case when $g_\rho=0$ and $g_\sigma~\sim 1$ is in the disordered phase. Since $\sigma$ is at $g_\sigma>1$, $\langle M_x\rangle_\sigma=0$ and ignoring the terms $C_\rho,C_\sigma$ in Eq.~\eqref{eq:main_eq_2} we get:
\begin{eqnarray}
\frac{D(\rho,\sigma)^2}{L^2}\approx\frac{1}{2}+\frac{1}{L}B_\sigma \tilde{g}_\sigma^{\nu(\eta-1)}.
\end{eqnarray}
We represent the extracted exponent $\nu(\eta-1)$ in Fig.~\ref{fig:fig3}(a) for different system sizes. As long as $L\gg \xi$, with $\xi$ being the correlation length, the power-law behavior of the subleading term is confirmed, obtaining $\nu(\eta-1)\sim -0.741$, fairly close to the exact value $\nu(\eta-1)=-3/4$. We understand this scaling as follows. In Eq.~\eqref{eq:main_eq_prev}, we integrate up to the correlation length $\xi$, which on the lattice yields $\sum_{n}\langle \sigma_i^x \sigma_{i+n}^x \rangle\sim\xi^{1-\eta}=\xi^{3/4}$. Now the correlation length $\xi$ scales as $\tilde{g}_\sigma^{-\nu}$ with $\nu=1$, which yields $\sum_n \langle \sigma_i^x \sigma_{i+n}^x \rangle\sim |1-g_\sigma|^{-3/4}$. A careful analysis of the formulas in Ref.~\cite{Pfeuty1970} shows that there are sub-leading corrections to the $n^{-1/4}$ dependency of correlations, which also influence the summation. However, the obtained result becomes exact in the thermodynamic limit, i.e. under Assumption 2. We stress here that the above result applies to the distance between different ground states that do not approach each other. Finally, in Fig.~\ref{fig:fig3}(b), we consider the case with $g_\sigma=10$, i.e. with the state $\sigma$ deep in the disordered phase. In that case, the leading contribution dominates the scaling, and one obtains the exponent $2\beta\sim 0.243$ to be fairly close to the analytic value $2\beta=1/4$ for the TFIM~\cite{Pfeuty1970}. 

To summarize, we have shown that the quantum Wasserstein distance exhibits clear signatures of criticality in many-body quantum systems hosting quantum critical points by revealing how it scales with system size and model parameters in both finite-size and thermodynamic-limit scenarios. We explicitly demonstrated this scaling in the integrable TFIM.

From a broader perspective, these findings offer new insights into recent approaches of learning many-body quantum states \cite{Carrasquilla2017,Bravyi2022,doi:10.1126/science.abk3333,Lewis2024,PhysRevResearch.6.033035} by tomographic and sampling based methods. Specifically the recent works \cite{Rouze2024quantum,Rouze2024} have employed the quantum Wasserstein distance of order 1 to efficiently infer expectation values of local observables with minimal sampling requirements. Importantly these methods require a "well behaved" phase of matter, i.e. exponentially decaying correlations. We note that the Wasserstein distance of order 2 has recently been used in a similar matter to learn quantum glassiness of ground states \cite{anschuetz2025efficientlearningimpliesquantum}. The generality of our results allows to investigate such methods and bounds now close to quantum criticality, in which  correlations decay only slowly with a power law. Addressing the modified scaling of the Wasserstein distance in these critical regimes could lead to refined sampling strategies that remain robust against long-range correlations. Similiar considerations apply to variational algorithms \cite{DePalma2023prxq,Zoratti2023,PhysRevResearch.6.043254,Fauseweh2023quantumcomputing} that aim to learn states with long-range correlations with minimal sampling overhead.

An interesting outlook is to consider the Wasserstein distance between two thermal states, i.e. arbitrary Gibbs states of the model at different temperatures, along the lines of Ref.~\cite{DePalma2023}. For the case of exactly solvable models, such as the TFIM, it is expected that even at finite temperatures, the complexity of the optimization problem in Eq.~\eqref{eq:wasserstein_distance} gets considerably reduced. A particularly interesting direction of future research is to investigate the scaling behavior of the quantum Wasserstein distance accross a quantum phase transition in non-equilibrium~\cite{Zurek2005}, as well as in dissipative systems~\cite{Iemini2018}. 

It is also interesting to explore whether solving the optimization problem in the set of separable couplings is enough to capture most of the critical behavior of this distance. This could have potential implications in finding efficient ways to discriminate distinct quantum states subjected to thermal equilibrium, but also to extract valuable information on the structure and the nature of entanglement in many-body systems displaying critical behavior.

\section*{Data availability}
The data supporting the findings of this study are available from Zenodo~\cite{zenodo_repo} and also upon request from the corresponding author.

\section*{Acknowledgments}
We would like to thank Géza Tóth for providing useful comments during the preparation of the manuscript. This project was made possible by the DLR Quantum Computing Initiative and the Federal Ministry for Economic Affairs and Climate Action of Germany; \url{qci.dlr.de/projects/ALQU}. Funded by the Deutsche Forschungsgemeinschaft (DFG, German Research Foundation) – Project number FA 1884/5-1.
\clearpage
\bibliography{bib}

\end{document}